**Graphical abstract**

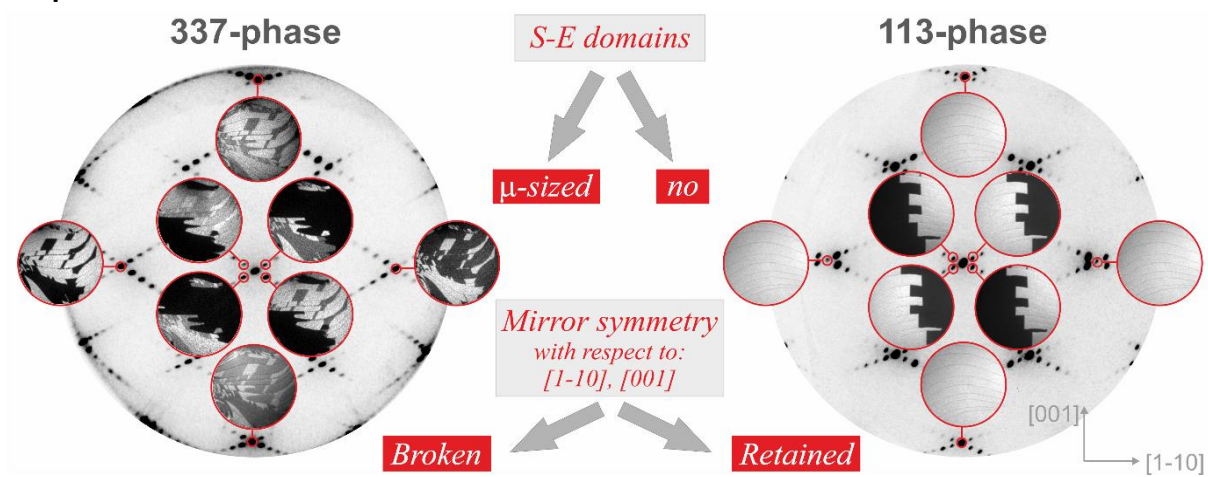

**Highlights:**

- Oxygen striped superstructures cannot be explained through alternating site-exchanged domains
- Superstructures originate from misfits between tungsten and oxygen lattices; Striped appearance corresponds to moiré pattern
- 113- and 337- phases differ in the average adsorption sites of oxygen atoms
- Thermal evolution of oxygen monolayer occurs via uniaxial reorganization along $\langle 1\bar{1}1 \rangle$ close-packed chains

# High-temperature oxygen monolayer structures on W(110) revisited

Dorota Wilgocka-Ślęzak[a,*], Tomasz Giela[b], Kinga Freindl[a], Nika Spiridis[a], Józef Korecki[a,c]

[a]*Jerzy Haber Institute of Catalysis and Surface Chemistry, Polish Academy of Sciences, ul. Niezapominajek 8, 30-239 Cracow, Poland*
[b]*National Synchrotron Radiation Centre SOLARIS, Jagiellonian University, ul. Czerwone Maki-98, 30-392 Cracow, Poland*
[c]*AGH University of Science and Technology, Faculty of Physics and Applied Computer Science, al. Mickiewicza 30, 30-059 Cracow, Poland*

*\*Corresponding author E-mail address: ncslezak@cyf-kr.edu.pl*

Systematic studies of the two high-temperature monolayer oxygen structures that exist on the (110) tungsten surface were performed using low-energy electron microscopy and diffraction measurements. Our work questions the commonly accepted interpretation from the literature that striped oxygen superstructures arise from alternating site-exchanged (S-E) domains. We postulate that the superstructures originate from a misfit between tungsten and oxygen lattices while the striped appearance corresponds to a moiré pattern.

Moreover, we show that the two structures, indicated as 113- and 337-phases due to the characteristic directions of the respective moiré patterns, differ considerably in their symmetry properties. This suggests that oxygen atoms in the two overlayers occupy different adsorption sites on average. In particular, the 113-phase features rotational domains that retain mirror symmetries with respect to the [001] and [1$\bar{1}$0] directions, whereas the 337-phase is characterized by the appearance of additional domains due to the breaking of these symmetries.

We propose structural models for both phases that consistently explain their unusual properties and suggest a universal mechanism for the thermal evolution of oxygen monolayer adsorbed on W(110).

## 1. Introduction

Two dimensional (2D) oxides on W(110) have been a subject of interest since the 1960s [1,2]. This simple system of a single oxygen monolayer chemisorbed on the tungsten surface and ordered at high temperatures has been investigated extensively for many years [3–14] as a special case of oxygen adsorption on metal surfaces. In parallel, oxygen adsorbed on W(110) with different sub-monolayer and monolayer coverages posed a model system for investigation of various aspects of adsorption-related phenomena, such as surface diffusion [15–21], kinetics of ordering 2D systems [22–24], interactions between adsorbed particles [25,26], surface phase transitions [27–32], structure of adsorption phases, [1,3,4,33], adsorption geometry [4,34], electronic [5,7,24,35] and vibrational [36,37] properties, surface stress [38–40], or the role of steps [41–43]. More recently, tungsten-oxide surfaces have attracted further interest as they can be readily utilized in electro-optical and semiconductor devices as catalysts or as gas sensors [44].

At moderate temperatures (<400 $^\circ$C) and sufficiently high oxygen partial pressures (>$1\times10^{-6}$ mbar), the W(110) surface adsorbs a (1x1) monolayer of oxygen with oxygen atoms occupying the pseudo-threefold hollow sites (h3) on the rhombic primitive unit cell of the tungsten surface [4]. Such adsorption positions break the symmetry with respect to the [001] axis, which causes site-exchange (S-E) domains to emerge as there are two equivalent adsorption sites within the unit cell. The presence of such domains, sized below 7.5 nm, was directly demonstrated using scanning tunneling microscopy (STM) by Johnson *et al.* [4] for an oxygen monolayer adsorbed at room temperature. The small domain size is likely to be crucial in relieving of the stress of the oxygen overlayer.

At elevated temperatures (above approximately 500 $^\circ$C), the stress of the (1x1) monolayer is reduced more effectively via the reorganization of oxygen atoms with slight desorption [4]. This process results in a characteristic spot superstructure in the LEED patterns [1–3,9,12–14], whereas the STM measurements show an array of highly ordered, several nanometer-wide stripes[1] with a local structure very close to (1x1) [4,12]. Additionally, the oxygen overlayer annealed above 1000 $^\circ$C under ultrahigh vacuum (UHV) conditions reveals there are uncovered tungsten rows between the stripes [4]. Rotational domains also appear as the stripes are aligned at an angle with respect to the high symmetry directions of the substrate. These domains usually span micrometer-sized areas with stripe directions and domain boundaries driven by the substrate step geometry [14]. Depending on the exact oxygen coverage, which is balanced around 1 ML, the oxygen overlayer can be organized in one of the two possible superstructures, which differ as the direction and width of the stripes [1,3,14]. For the phase that is slightly more oxygen rich, denoted here as the 337-phase, 25-Å wide stripes run along the [$3\bar{3}7$] or [$\bar{3}37$] directions [12,14]. Whereas in the case of the 113-phase, with a lower oxygen coverage, the stripes are oriented along the [$1\bar{1}3$] and [$\bar{1}13$] directions [4] and are approximately 30-Å wide [14].

Most of the studies on such structures are focused on the 337-phase [1,3,6,9–14], whereas the 113-phase has been much less examined [1,3,4,14]. It has been experimentally demonstrated that for the 337-phase, the oxygen adsorption site is located at or near the pseudo-threefold h3 position [6,9,11], where an oxygen adatom has three equal bond distances. For the 113-phase, the adsorption site has not been experimentally determined and was instead assumed *a priori* as h3 [4]. It is noted that theoretical calculations suggest that a small (approximately 0.2 eV) energetic barrier between the two energy minima of the twin h3 sites can be further lowered for some specific configurations of neighboring oxygen atoms, making the twofold site between the twin h3 hollows (long bridge – lb) energetically favorable for adsorption [31]. Therefore, for the 113-phase, an alternative oxygen adsorption position should be considered.

For both phases, the recurrent explanation for the nature of the stripes in the literature involves the S-E phenomenon, for which adsorption at the twin h3 positions is the prerequisite. This explanation was based on the observation by Johnson *et al.* [4] that the (1x1) oxygen monolayer, which appears as a network of small S-E domains when adsorbed at room temperature, transforms into the 113-phase striped superstructure via UHV annealing. They proposed a model in which the network of the S-E domains organizes into alternating nano-sized stripes with the h3 adsorption site of oxygen switching from stripe to stripe between the two mirror-symmetric locations. This idea combines the S-E domains (*a priori* assumption of oxygen adsorption at the h3 position) and the striped superstructure into one phenomenon, in which the stripes are made from the S-E domains with a locally ideal (1x1) structure. This interpretation of stripes in the 113-phase has been extended by other authors to the second type of oxygen superstructure, the 337-phase [6,9,10], and the model became a common explanation for all observed monolayer oxygen superstructures on W(110), although the superstructure stripes have never been directly validated as S-E domains. In parallel, the alternative interpretation involving lattice misfits that leads to a coincidence superstructure, as proposed by Bauer and Engel [3], has largely been abandoned. However, more recent studies based on photoelectron holography suggest that the oxygen monolayer on W(110) surfaces in the 337-phase does not form the perfect (1x1) structure (within individual stripes) but forms a slightly expanded one instead [11]. Similar conclusions regarding deviations in the oxygen layer from the ideal (1x1) arrangement can also be derived from studies of atomically resolved STM [12].

---

[1] The term *stripes* will be used throughout the paper to describe linear periodic structures observed via STM. The nature of these stripes itself is a subject of the work.

Another open issue is the exact adsorption position of oxygen in the 337-phase. There is a body of evidence from photoelectron diffraction experiments, which measure the local symmetry of the emitter, that oxygen atoms in the 337-phase overlayer do not sit at the h3-site, but shift along the $[1\bar{1}0]$ direction [6,9]. On the other hand, Menteş and Locatelli [13] observed an additional contrast level in the domain structure of the 337-phase using dark-field low-energy electron microscopy (df-LEEM) and dark-field X-ray photoemission electron microscopy (df-XPEEM), which was interpreted as originating from the broken mirror symmetry with respect to the $[1\bar{1}0]$ axis, i.e., due to a small shift in the oxygen adsorption position from the h3 position along the $[001]$ direction. The mutually orthogonal shifts reported in the two experiments suggest that the displacement of the oxygen atoms from the h3 position in the 337-phase has much more complex characteristics.

From the above discussion, fundamental questions arise concerning the real nature of high-temperature oxygen striped superstructures on W(110). Does the oxygen lattice form a locally perfect (1x1) structure with alternating S-E stripes, or is the superstructure due to the coincidence between a strained oxygen lattice and the tungsten surface mesh? What is the adsorption site for the oxygen atoms and is it identical for the two different oxygen structures? Which mechanism governs the thermal evolution of the oxygen overlayer?

To answer these questions, we performed a comparative study of the 337- and 113-phases using df-LEEM imaging combined with micro diffraction LEED (μ-LEED) I-V curve analysis. The analysis led to structural models for the two oxygen structures and allowed to propose a universal mechanism for the thermally induced transformations of the oxygen monolayer on W(110).

## 2. Materials and methods

The experiments were conducted in an ELMITEC LEEM III system with a base pressure of $5 \times 10^{-11}$ Torr. The 337-phase was obtained via exposing a clean W(110) single-crystal (cleaning procedure we described in [14]) to a 1x10⁻⁶ Torr molecular oxygen atmosphere at a temperature of approximately 1000 °C for 10 min. The UHV annealing above 850 °C transformed the 337-phase into the 113-phase within a few minutes. Alternatively, the oxygen superstructures were produced by UHV annealing of the (1x1) oxygen monolayer adsorbed at 1x10⁻⁶ Torr below 300 °C. The 337-phase emerged above 450 °C and transformed into the 113-phase when above 850 °C. The development of the phases was monitored in real time using both LEED and LEEM. It is noted that the temperatures of the transformations strongly depend on a local step geometry of the W(110) surface.

The structural properties of the high-temperature oxygen phases were scrutinized using df-LEEM imaging and μ-LEED measurements as functions of the electron energy. The superstructure stripes could not be directly observed from real-space imaging as their widths were below the resolution limit; thus, the LEEM analysis was based on imaging the micro-sized structural domains arising from surface symmetry breaking. The comparative μ-LEED I-V analysis provided the symmetry properties for these domains.

Typical LEED patterns collected on multi-domain areas of the 337- and 133-phases are shown in Figures 1a and 2a, respectively. Characteristic crosses formed from superstructure spots centered around the indexed tungsten spot result from the overlapping diffracted beams that originate from areas (rotational domains) revealing mirror reflected stripe superstructures. The red circles in the LEED patterns mark the spots used for the domain structure imaging via df-LEEM. The imaging was performed for each phase in three series by exploiting three types of diffraction spots selected with a small contrast aperture: *series I* – the 1st order satellite beams around the specular beam (indexed from 1 to 4); *series II* – the (1,1) and (-1,1) spots; and *series III* – the (1 -1) and (-1 1) spots. The three series of the df-LEEM images for the 337- and 113-phases are shown in Figures 1b–d and 2b–c, respectively. The images are labeled with the corresponding spot indexes. Only a single representative image is shown when the acquired images with different spots revealed the same contrasts. The imaging was complemented with a series of μ-LEED patterns collected as functions of the electron energy from all types of domains resolved in the df-LEEM images. Finally, the I-V curves were derived from these μ-LEED series.

All measurements were performed at room temperature. All the LEEM images and LEED patterns, as well as the models for the surface structures are oriented with the [1$\overline{1}$0] direction running horizontally.

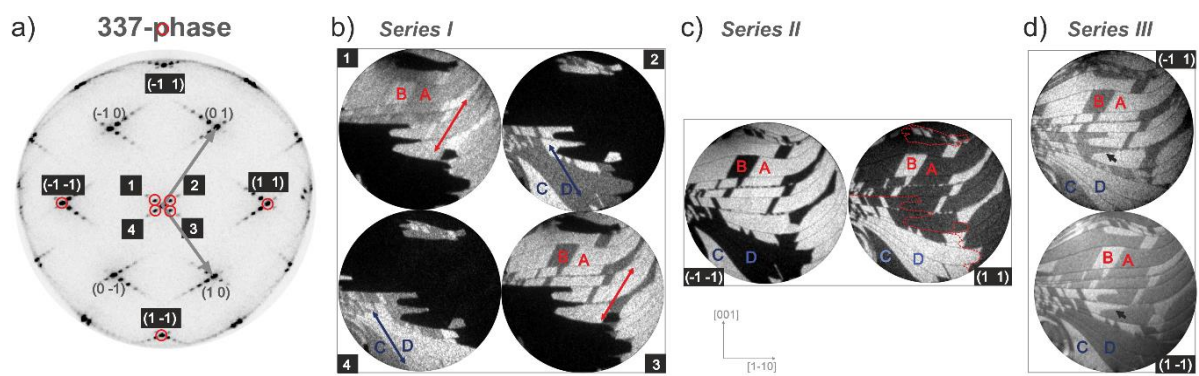

Figure 1. LEED pattern (a) and df-LEEM images (b–d) collected for the 337- phase. The df-LEEM images were collected in series by exploiting three different types of diffraction spots (marked by red circles and labeled on the diffraction pattern; gray vectors on the pattern describe the primitive unit cell for tungsten): b) '*series I*' – images taken with 1–4 superstructure spots, c) '*series II*' – with (11) and (-1-1) spots, and d) '*series III*' – with (1-1) and (-11) spots. The field of view (FoV) is 5 um, and the red and blue bidirectional arrows depict the actual directions

of the stripes in the highlighted regions; A, B, C, and D denote the different sub-domains; and the dashed red line in the (1 1)-image represents the border between areas with different stripe directions. All data were acquired at an electron energy of 105 eV.

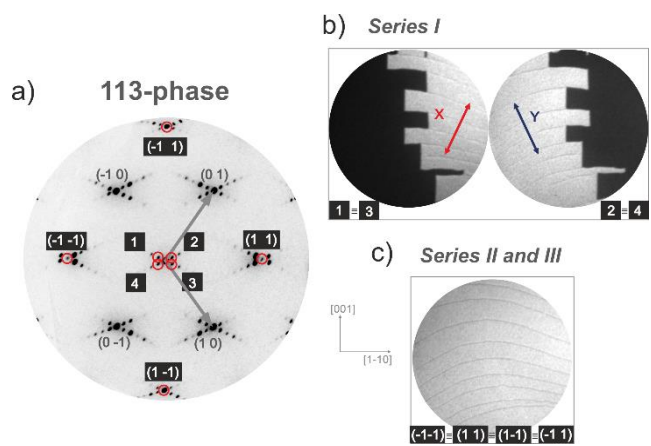

Figure 2. LEED pattern (a) and df-LEEM images (b, c) collected for the 113- phase. The df-LEEM images were collected in series by exploiting three different types of diffraction spots (marked by red circles and labeled on the diffraction patterns; gray vectors on the pattern describe the primitive unit cell of tungsten). b) '*series I*' – representative images for 1-4 superstructure spots, c) '*series II and III*' – representative image for (11), (-1-1), (1-1) and (-11) spots. FoV=5 um. The red and blue arrows depict the actual directions of stripes in the highlighted regions; X and Y label different rotational domains. All data were acquired at an electron energy of 105 eV.

## 3. Results and analysis

### 3.1. 337-phase

#### 3.1.1. Series I

The domain structure of the 337-phase as imaged using the superstructure spots is shown in Figure 1b. It was previously demonstrated by Menteş and Locatelli [13] that the strong contrast corresponding to the rotational domains associated with the direction of the stripes further splits into sub-contrasts to disclose another system of domains. In Figure 1b, these sub-domains are labeled as A and B as well as C and D on the odd- and even-spot images, respectively. For the A and B sub-domains, the stripes run along the same direction (marked with red bidirectional arrows oriented along $[3\bar{3}7]$), whereas the C and D sub-domains exhibit mirror reflected $[\bar{3}37]$ rows as indicated with blue arrows. The sub-domain contrast is swapped between images acquired with the satellite spots that belong to the same rotational domain (1 and 3 or 2 and 4). Based on the interpretation by Menteş and Locatelli [13], the sub-contrast originates from the symmetry breaking along the [001] direction (i.e., with respect to the $[1\bar{1}0]$ axis), while the striped superstructure responsible for the main contrast is linked to the presence of two equivalent triply-coordinated adsorption sites, which is consistent with Johnson *et al.*'s model [4]. We propose an alternative explanation for the observed domain structure as supported further in the *series II* results (described in Section 3.1.2.), which attributes the sublevel contrasts to large S-E domains and thus excludes the alternating ordered S-E domains as the possible origin of the stripes.

#### 3.1.2. Series II

The images obtained with the (11) and (-1-1) spots for the 337-phase (Figure 1c) reveal a simple two-level contrast, which perfectly reproduces the shapes of the sub-domains seen from *series I*. Of note, the correlations between the contrasts observed by Menteş and Locatelli [13] in the df-LEEM images collected on the superstructure spots and in the df-XPEEM image, which is sensitive to the local symmetry of the oxygen atoms, are identical to the correlations seen in our images from *series I* and *II*. This means that the df-XPEEM and our df-LEEM images for *series II* show the same type of domains, where the A and C sub-domains (with mutually mirrored directions of stripes) merge into one contrast level, whereas B and D (representing opposite rotational domains) contribute to the opposing contrast. Our μ-LEED I-V curves measured on the (11) and (-1-1) spots for the four single sub-domains (Figure 3a) clearly indicate that diffractions along the two opposite directions of $[1\bar{1}0]$ (i.e., rightward and leftward) are nonequivalent as the two corresponding I-V curves are distinctly different for each individual sub-domain. In parallel, all the domains are represented by qualitatively similar sets of the two curves that swap their shapes for the (11) and (-1-1) spots, with the most characteristic feature being the low energy maximum that switches between 40 and 50 eV.

An analysis of the interrelations for the curve behaviors between the sub-domains indicates that A, C and B, D pose two symmetry equivalent pairs of sub-domains, which are mutually mirror related, i.e., the adatom arrangement with respect to the [001] axis in the two pairs are mirror reflected. Similar conclusions are derived from the analysis of the I-V curves collected at the {10} spots. Figures 3b and c juxtapose the curves acquired for the spots located on the right (R) and left (L) sides of the diffraction pattern, respectively. The curves collected for the given domain with the L or R spots are qualitatively very similar, and in parallel, the L and R curves are completely different. Thus, it is clear that the dominating contribution to the observed A, B, C, and D sub-domain contrasts originates from the broken mirror symmetry with respect to the [001] axis instead of the $[1\bar{1}0]$ as suggested in Ref. [13]. The symmetry breaking with respect to the $[1\bar{1}0]$ axis and its impact on the domain structure is discussed in Section 3.1.3 based on *series III*.

Photoemission diffraction experiments on the 337-phase [6,9] indicate that the broken symmetry for the local arrangement of oxygen atoms with respect to the [001] axis in this system is because they occupy the h3-type hollows, which reveal a lowered symmetry. Moreover, these data show that the adatoms are slightly shifted (approximately 0.1 Å) along the $[1\bar{1}0]$ from their exact h3 positions [6,9], and which is always outwards from the tungsten primitive cell center.

This combination of the experimental data implies that the observed micro-sized sub-domain system is due to the S-E effect as a result of the two mirror h3 adsorption sites. This interpretation is additionally confirmed from a pseudo-triple symmetry reflected in the LEED intensities, as verified by the intensity analysis of the selected (1x1) diffraction spots for individual sub-domains as functions of the electron energy. We found that for some energy ranges (this effect is especially pronounced between 35 and 55 eV), the spots form two triplets with strongly enhanced and reduced intensities, as shown in Figure 3d. As expected, the LEED patterns from the A and C sub-domains reveal the same pseudo-triple symmetry, whereas those measured for B and D sub-domains are mirror reflected, which clearly demonstrates the pseudo-triple symmetry is related to the symmetry of the adsorption sites. It is noted that the explanation of the micro-sized sub-domains detected using df-XPEEM and df-LEEM in terms of the S-E domains, in which oxygen occupies the alternative pseudo triply-coordinated adsorption positions, requires a new interpretation for the nano-sized stripe-structure [4], as discussed in the following sections.

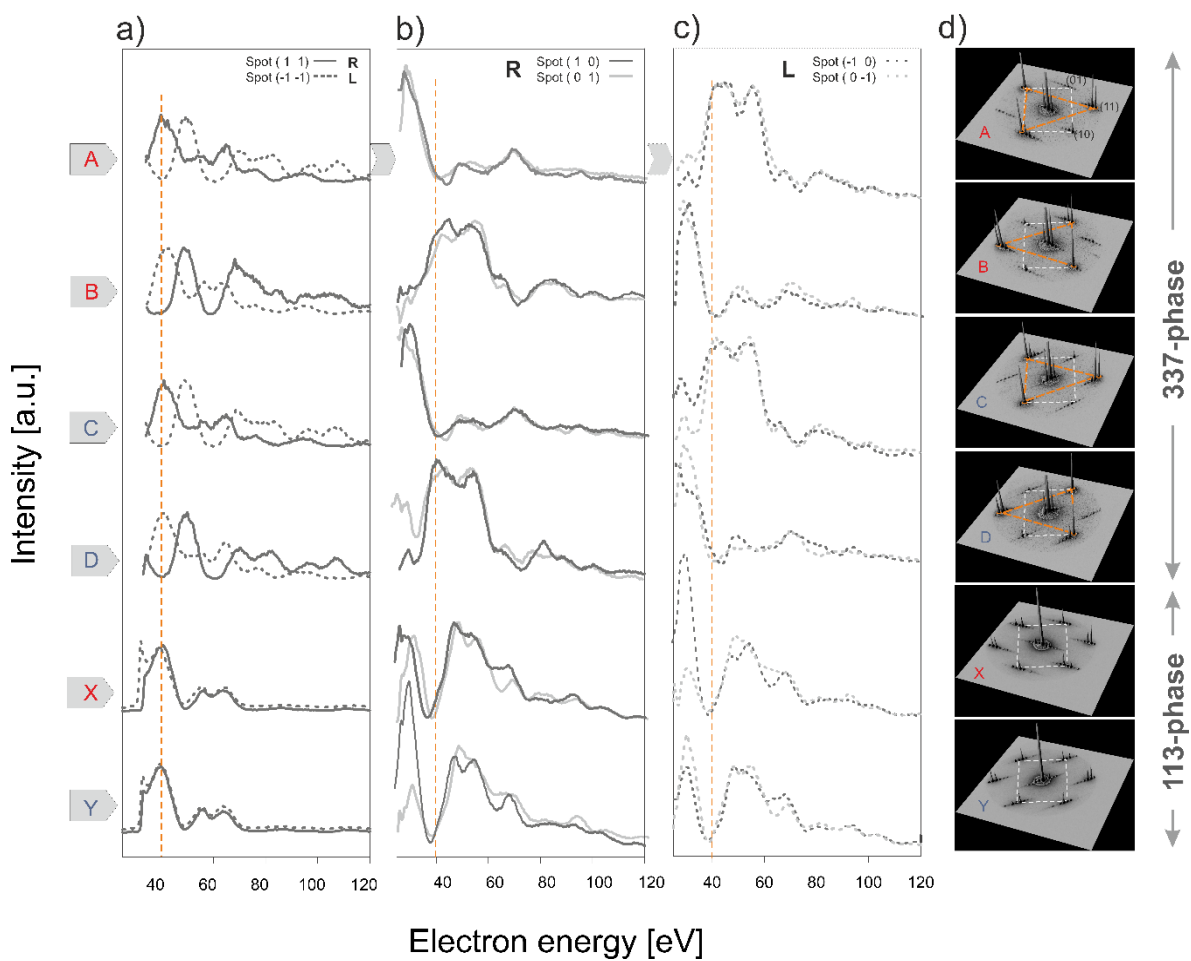

Figure 3. LEED I-V curves collected from individual A, B, C, and D sub-domains (337-phase) and X and Y (113-phase) domains for a) (-1-1) and (11) spots, b) (0 1) and (1 0), and c) (-1 0) and (0-1) spots (L and R denote the left and right sided spots of the diffraction pattern, respectively). d) The 'intensity landscapes' of the LEED patterns collected at an electron energy of 40 eV (energy marked on the (a)-(c) I-V curves as dashed orange lines. Dashed lines join the spots corresponding to W(1x1): the {10} spots are connected as the white line whereas the orange line links the most intensive spots out of {10} and {11}.

### 3.1.3    Series III

The 337-phase observed with the (1-1) and (-11) spots (Figure 1d) reveals a domain structure with a simple two-level contrast,[2] which suggests symmetry breaking with respect to the [1$\bar{1}$0] axis. That is, there is an additional displacement component of the oxygen atoms in the [001] direction. The domain boundaries are combinations of those observed in *series I* and *II* as they reflect the shapes of both the S-E domains and the rotational domains (e.g., compare the domain wall indicated with the black arrow that separates the A- and C-type domains, which belong to the same S-E domain). The contrast in the domains imaged for this series shows specific interrelations with the contrasts obtained for the corresponding areas in the other series, as schematically shown in Figure 4a. The figure illustrates the experimental contrasts from Figure 1b–d with the corresponding shading of the squares that represent the A, B, C, and D sub-domains. Based on the known actual directions of stripes and the relationships between the occupied adsorption hollows in the A, B, C, and D sub-domains, we arranged them into an array where the stripes and the adsorption positions of neighboring domains are mirror reflected, as shown in Figure 4b (please ignore the black and light blue arrows) . For such domain arrangements, the corresponding experimental contrasts form patterns with a characteristic symmetry for a given series. Specifically, the contrast patterns for *series I* and *II* show expected 'diagonal' and 'horizontal' symmetries, respectively, which reflect the properties of the stripes and site-exchange phenomena (as well as the [1$\bar{1}$0]-shift). Surprisingly, the pattern for *series III* shows a 'vertical' symmetry, which indicates that the [100]-shift of the adatoms is correlated with the direction of the stripes and adsorption position. Namely, the contrasts indicate that the A and D sub-domains (C and B) that represent mirror-reflected stripes and alternative adsorption sites exhibit the same [001]-shifts for oxygen from the h3 position. This indicates that the direction of stripes together with the type of the adsorption position determines the up- or down- direction of the [001]-shift.

---

[2]A weak contrast is noticeable on one of the images as an unwanted contribution that originates from the superstructure spots around (1-1) and (-11); it was difficult to completely avoid these components as the images were collected under critical conditions – very faint intensity and strongly bent electron beams.

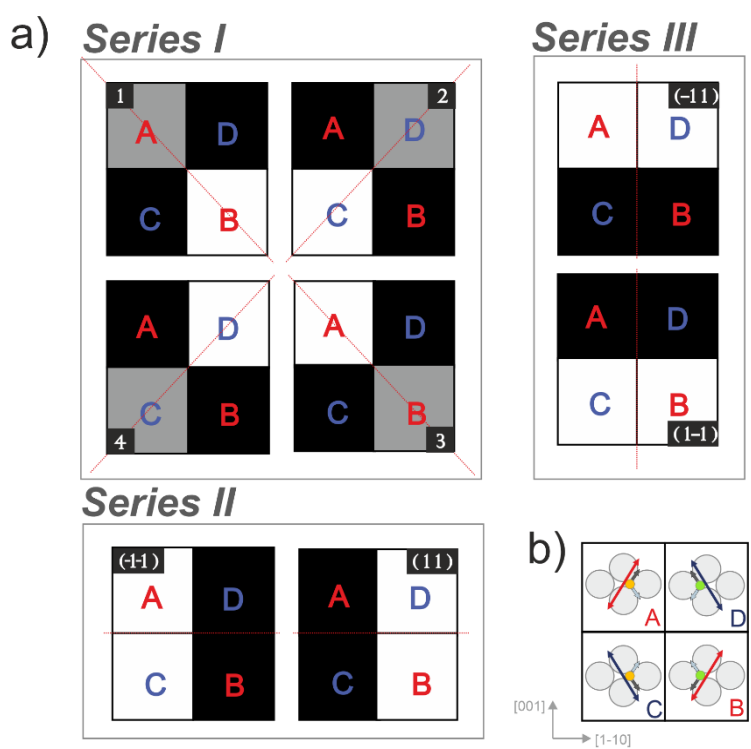

Figure 4 a) Experimental contrasts of the A, B, C, and D sub-domains for *series I, II,* and *III* as derived from df-LEEM images of Figures 1b–d. The domains are represented with squares arranged into an array as shown in the pictogram b): blue and red bidirectional arrows mark the actual direction of the stripes, while green and yellow circles within the W-unit cell (gray circles) depict the adsorption positions. The black and light blue arrows represent two sets of exemplary shifts of oxygen atoms, which could explain the contrast symmetry for *series III* (it was assumed that oxygen moves along the paths located between tungsten atoms). The red dotted lines in a) correspond to the respective symmetry axes of the contrast patterns.

The experimentally demonstrated presence of two mutually orthogonal directions in the shifts (i.e., [1$\bar{1}$0] and [001]) indicates that they represent components of shifts that occur along certain higher-index directions. The photoelectron diffraction results show that the horizontal component of the shift is directly correlated with the occupied h3 hollow (left or right),  and is always outwards from the center of the W unit cell [6,9], whereas the vertical component (down or up) is determined from two effects: the actual h3 position and the stripe direction. In Figure 4b, two exemplary sets of four possible directions of the shift, consistent with the experimental data, are depicted by black and light-blue arrows. The next section proposes a model to explain the origin of the stripes and answers the question of what is the reason that the shift is strictly correlated with the adsorption h3 site and the stripe direction.

We note that the df-XPEEM image published by Menteş and Locatelli [13] reveals a faint, unreported sub-contrast in addition to the main contrast, which is consistent with our *series III* and gives identical correlations to the boundaries of the rotational and S-E domains. Based on our present interpretation, it is concluded that the main contrast on the df-XPEEM image is related to the S-E domains, whereas the weak sub-domains correspond to the atomic oxygen displacements along the four equivalent directions.

### 3.1.4. Model for 337 structure

The stripes in the 337-phase cannot be explained from alternating site-exchanged domains. An alternative explanation of the striped superstructure, proposed by Bauer and Engel [3], invokes a misfit between the tungsten surface lattice and the oxygen overlayer which leads to a coincidence structure; in this case the characteristic multi-spot LEED pattern is understood as a result of double scattering between the two lattices. In addition, the photoelectron holography results from Takagi *et al.* [11] suggest that the oxygen overlayer is slightly expanded with respect to the substrate, whereas the atomically resolved STM image of the 337-phase [12] indicates that close-packed oxygen rows are declined from the $W[1\bar{1}1]$ direction by approximately 2.5°. If there is misfit between the oxygen and tungsten lattices, it would be reflected in the LEED pattern as the electrons that experience only single scattering within the oxygen overlayer should intensify specific spots with respect to the other superstructure spots. Therefore, within the kinematic approach, we analyzed the spot intensity for series of LEED images collected over a broad energy range for individual sub-domains. Figure 5a shows the behavior of the spot profiles across the W(110)-(1x1) and surrounding satellites as a function of energy between 40 and 120 eV for the A sub-domain. The maxima corresponding to the tungsten spots are marked with gray arrows. Figure 5b shows the LEED patterns with profiles averaged over the entire energy range, which represent the typical behavior of the intensity relationships between spots, for the A and D sub-domains.

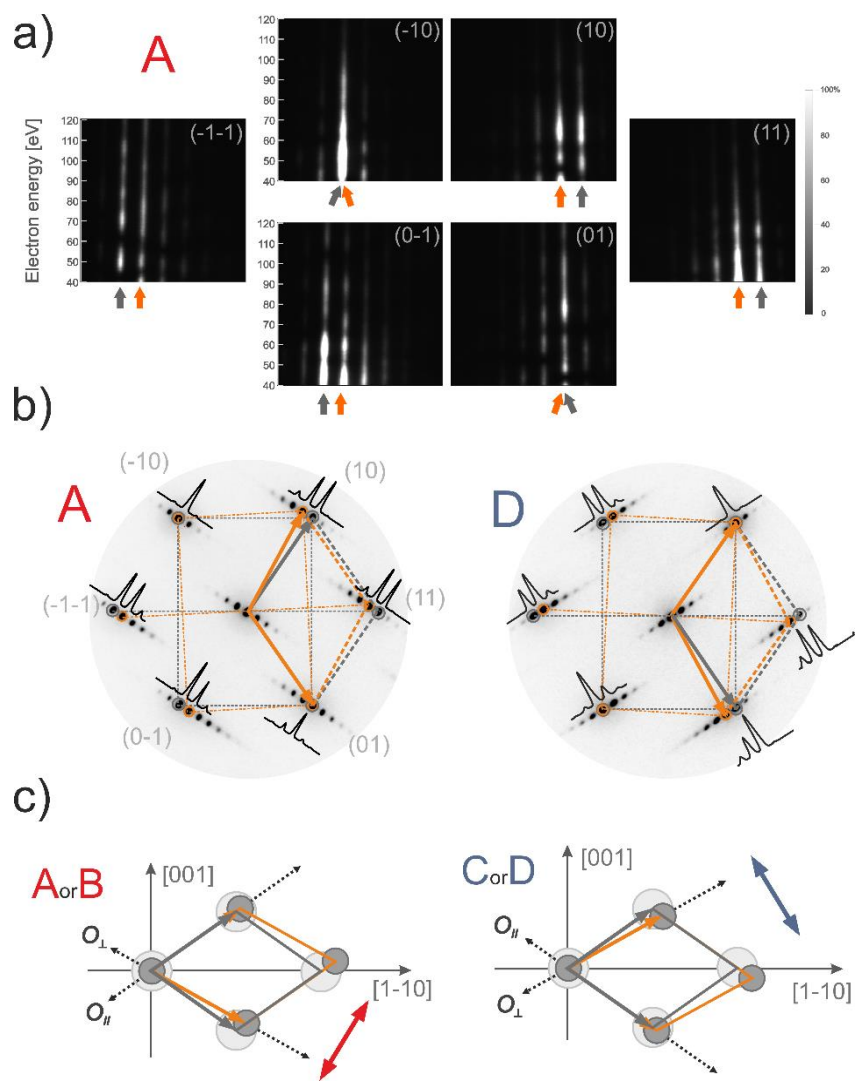

Figure 5. Data derived from the LEED patterns collected for the 337-phase. a) Changes in the intensity profiles across the W(1x1) spots and their satellites as a function of energy (40–120 eV) on sub-domain A. The positions of the tungsten-related spots are indicated with gray arrows; the maxima ascribed to the oxygen overlayer are marked with the orange arrows. b) Energy-averaged (40–120 eV) LEED patterns and intensity profiles (normalized) that reflect the typical relationships between spot intensities for sub-domains A and D; gray and orange coded circles and vectors mark the spots and unit cells corresponding to tungsten and oxygen lattices, respectively. c) The unit cells in the real space corresponding to sub-domains A or B and C or D (bigger and smaller circles represent the W and O atoms, respectively) in the 'on-top' configuration to highlight the relationship between the two cells. The red and blue bidirectional arrows depict the directions of the stripes for the structures.

For each domain, four of the six visible W(110)-(1x1) spots marked with gray circles in Figure 5b (or gray arrows in Figure 5a) have a satellite with a comparable or even higher intensity (closest to the main spot), which dominates the other satellites. The two-remaining diagonal {10} spots clearly dominate their satellites with one of them being much stronger than the other W (1x1) spots (compare with Figure 5a). This suggests that the substrate and overlayer in this direction have the same periodicity and diffraction spot overlap. Thus, the four intense satellites and two (1x1) spots, marked by orange circles (and orange arrows in Figure 5a), are identified as the reciprocal lattice nodes of the oxygen overlayer. The assumed oblique oxygen and rhombic tungsten unit cells in real space are shown in Figure 5c in orange and gray, respectively. For the oxygen structure, one of the close-packed atomic directions follows the direction of the corresponding close-packed tungsten row (denoted as $O_{||}$), whereas the other one is rotated by approximately 6° from the corresponding dense-packed tungsten row. To model the coincidence structure, it is important to relate the oxygen close-packed directions to the direction of the superstructure stripes, which is closer to the $O_{||}$ and nearly perpendicular to the second oxygen close-packed direction, denoted as $\mathbf{O}_\perp$. To reconstruct the coincidence structure, we additionally consider the atomically resolved STM literature data that shows the 337-phase has nine oxygen atomic rows within a stripe and a 5-atom periodicity along the stripes [12]. Testing different possible coincidence superstructures, we found that only one gives the ~6° misalignment between the oxygen and tungsten $\langle 1\bar{1}1 \rangle$ chains,[3] which can be described in matrix notation (expressed with a surface tungsten primitive base with a spacing of u=2.73 Å) as:

$\mathbf{G_{coinc}}=\begin{bmatrix} 9 & 1 \\ -2 & 5 \end{bmatrix}$ with the oxygen lattice given by $\mathrm{G_O}=\begin{bmatrix} 1 & \frac{1}{9} \\ 0 & \frac{47}{45} \end{bmatrix}$, for the A and B sub-domains

$\mathbf{G_{coinc}}=\begin{bmatrix} 5 & -2 \\ 1 & 9 \end{bmatrix}$ with the oxygen lattice given by $\mathrm{G_O}=\begin{bmatrix} \frac{47}{45} & 0 \\ \frac{1}{9} & 1 \end{bmatrix}$, for the C and D sub-domains

The above superstructure for A sub-domain is visualized in the Supplementary Information in Figure S1a. The coincidence-cell contains 47 W vs. 9x5=45 O atoms. This means that there are 4.25% fewer oxygen than tungsten atoms, indicating the oxygen lattice is expanded[4] . In this structure, the $\mathbf{O}_\perp$ rows are rotated by 5.77° against the close-packed W rows. The distance between oxygen atoms along these chains is 1.0423u. For the $\mathbf{O}_{\parallel}$ rows, which exactly trace the complementary W$\langle 1\bar{1}1 \rangle$ chains, the periodicity of the oxygen atoms is 1.04(4)u. The angle between the two oxygen rows is 64.76° and the lattice loses its mirror symmetry with respect to the $[1\bar{1}0]$ and [001] tungsten directions. The parameters of this structure are collected in Table 1. A closer inspection of this model (for details, compare Figure S1a) shows that the observed deformation of the oxygen lattice with respect to the (1x1) structure is a consequence of a simple reorganization of oxygen atoms along the $\mathbf{O}_{\parallel}$ chains: 45 oxygen atoms match to 47 tungsten atoms and the successive expanded oxygen chains slip mutually by (1/9)u. In other words, the oxygen overlayer can be imagined as an array of oxygen closed-packed chains with a reduced density of atoms, which are parallel to one of the closed-packed directions of the substrate retaining the same periodicity between the chains as in tungsten. This unidirectional reorganization within and between the close-packed oxygen rows results in a moiré striped superstructure with the coincidence stripes along one of the two mirror $\langle 3\bar{3}7 \rangle$ directions (along the stripe direction, specific oxygen lines occupy the equivalent local adsorption sites).

In our model of the coincidence structure, the adsorption position is understood as the average position of the oxygen atoms in the coincidence unit cell. Thus, the average position should reflect the left or right h3 adsorption site (including the $[1\bar{1}0]$ shift from the ideal h3 position), as well as the oxygen displacement along [001], which is necessary to explain the contrast observed in *series III*. In order to create an appropriate model, we first analyzed the coincidence structure with the ideal average $\overline{h3}$ site. All possible combinations of the ideal $\overline{h3}$ sites and the directions of stripes give four domains (as shown in the Supplementary Information in Figure S2). However, they would not provide expected for the *series III* contrast due to preserved mirror symmetry of the average adsorption site with respect to the $[1\bar{1}0]$ direction. Moreover, the ideal $\overline{h3}$ position does not appear to provide the best match for the oxygen overlayer to the tungsten surface as the $\mathbf{O}_{\parallel}$ chains, which exhibit the same periodicity as the corresponding $\langle 1\bar{1}1 \rangle$ close-packed tungsten rows, are not centered between the W rows but are displaced from these grooves by 0.32 Å (details are included in the Supplementary Information in Figure S3).

---

[3]Superstructure proposed in the literature $\begin{bmatrix} 9 & 0 \\ -2 & 5 \end{bmatrix}$ [12,14] is inconsistent with the interpretation of the stripes as a result of the coincidence O(1x1)-W(110) structure.

[4]As a comparison, the superstructure proposed by Bauer and Engel [3] was selected with alternative satellite-spots with a ratio of 1.08 for the oxygen to tungsten atoms; giving a compressed oxygen overlayer.

To achieve this 'centered' position, which intuitively appears to be the optimal one, the oxygen overlayer has to be specifically displaced relative to the tungsten layer. Assuming that the $[1\bar{1}0]$-shift equals 0.1 Å out from the tungsten unit cell from the ideal $\overline{h3}$ position [6,9], i.e. it is rightward for the A and C and leftward for B and D sub-domains, the [001]-shift should be 0.46 Å up for the A and D and down for B and C sub-domains. This naturally leads to four unique shifts, which are driven by the symmetry of the specific h3 hollow and the direction of the unidirectional reorganization (thus stripes). The four coincidence structures in the proposed model are shown in Figure 6 with directions for the shifts indicated. It is apparent that the real atomic structure is likely not as simple as proposed here, and additional vertical and/or lateral modulation of the local positions of oxygen atoms in the coincidence cell should be considered. To validate the model and determine all details theoretical calculations are indispensable, however the big coincidence unit cell is challenging. Nonetheless, it is noted that the above model provides the mechanism for symmetry breaking, explains the striped superstructure, reproduces all the contrasts and their interrelations observed in our study, and coalesces all the diverse results from the literature into one coherent picture.

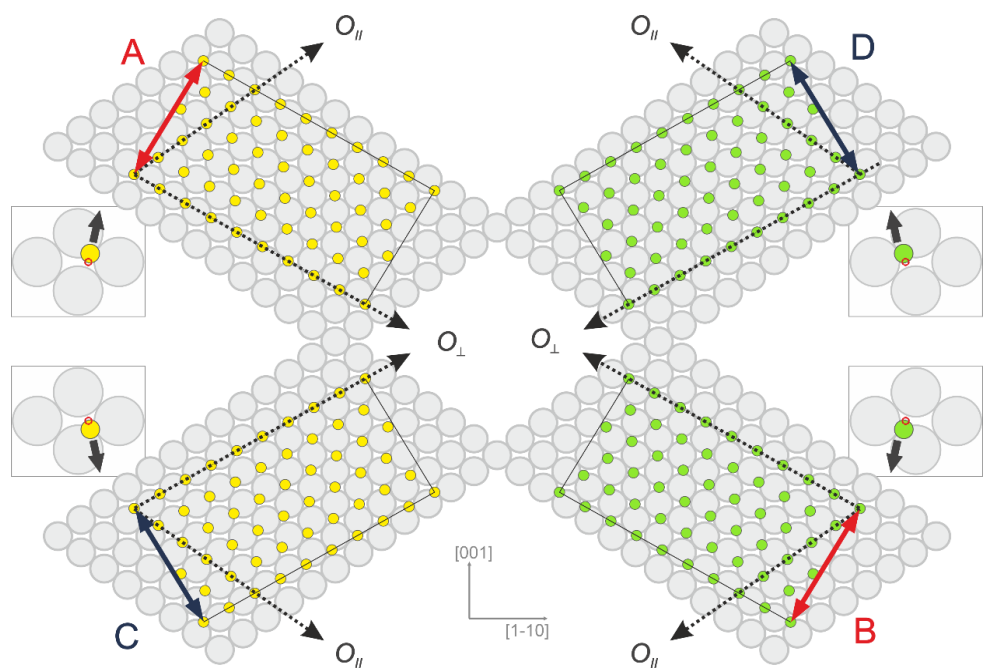

Figure 6. Coincidence unit cells for the A, B, C, and D sub-domains in the 337-phase with modified $\overline{h3}$-type adsorption sites. Larger circles represent the (110) tungsten surface, while smaller ones correspond to the oxygen lattice (described by $\mathbf{G_O}$ matrixes - see main text). The W and O lattices form the two coincidence structures (described by $\mathbf{G_{coinc.}}$ matrixes) that differ in the stripe directions (red and blue bidirectional arrows). Left and right h3-type adsorption positions (green and yellow colors) modified with unique shifts (that provide an optimal match between the O and W lattices, i.e., center $\mathbf{O_{\parallel}}$ chains in the grooves between close-packed tungsten rows) give rise to four domains with broken mirror symmetries with respect to the [001] and $[1\bar{1}0]$ axes. The close-packed oxygen rows $\mathbf{O_{\perp}}$ and $\mathbf{O_{\parallel}}$ are marked with dashed arrows. Pictograms in the insets show the average adsorption positions within the coincidence unit cells, which are shifted by 0.1 Å horizontally and 0.46 Å vertically from the ideal $\overline{h3}$ positions (marked with small red rings; shifts are depicted with thick black arrows). The model reproduces all the properties of the contrasts observed in the df-LEEM and LEED measurements.

### 3.2. 113-phase

### 3.2.1. Series I-III

The images collected using the superstructure spots for the 113-phase reveal a simple two-level contrast due to the rotational domains with the mirror-reflected directions of the stripes, as shown in Figure 2b. The images as obtained with different spots from the same domain are identical (1 ≡ 3 and 2 ≡ 4). The red and blue arrows indicate the actual directions of the stripes in the bright X and Y domains for $[1\bar{1}3]$ and $[\bar{1}13]$, respectively. The other two series, II and III (Figure 2c), of the dark field images do not reveal domain-contrast and only atomic tungsten steps are visible, which indicates that, in a mesoscale, the 113-phase is symmetric in both the $[1\bar{1}0]$ and $[001]$ directions.

The μ-LEED I-V curves acquired for the W(1x1) spots of the X and Y domains are identical (Figure 3a) or have very similar characteristics (Figures 3b and c). Noticeable differences may arise from differences in the diffraction along and across the stripes, and there may be deviations from the perfect perpendicular incidence of the primary electron beam. Furthermore, the pseudo-triple symmetric behavior of the selected W(1x1) spots observed for the 337-phase is no longer present in the 113-phase (Figure 3d).

From the above experimental observations, it is clear that mesoscale S-E domains of the 337-phase characteristics do not exist in the 113-phase. On the other hand, the lack of contrast does not exclude alternating striped S-E domains that are consistent with Johnson et al.'s model [4]; however, the existence of the striped S-E domains is highly improbable due to the reversible transitions found between the two phases [14]. The transitions are driven by temperature-dependent desorption/adsorption of oxygen in a temperature range 700–1100 °C for heating/cooling under an oxygen atmosphere between $5x10^{-8}$–$5x10^{-7}$ Torr, with the 113-phase nucleating via heating. The observed reorganization from the 337-phase into the 113-phase is accompanied with vanishing S-E domains, which reappear when the transition occurs in reverse direction. Clearly, the nature of both phases is similar, especially that there is a temperature range where both phases coexist. Therefore, the explanation of stripes in terms of nanometer-sized alternating S-E domains with a local (1x1) oxygen structure also fails for the 113-phase. Logically, the 113-phase should also be a coincidence structure but with different average adsorption position of oxygen because the strong manifested low-symmetry in the average adsorption positions of the 337-phase is lost in the 113-phase. Consequently, the average adsorption position should be highly symmetric at the W(110) surface, i.e., the long bridge position ($\overline{bl}$), as foreseen by theory [31].

### 3.2.2 Model for 133 structure

The structural model for the 113-phase was based on the assumption that the structure retains the characteristic features of the 337-phase, and that the oxygen lattice is represented in diffraction patterns with the same satellites around the W(1x1) spots. Figure 7a shows the averaged energy (40–70 eV) LEED pattern with intensity profiles for the X domain, where the most intensive satellites, closest to the tungsten spots, are ascribed to the oxygen overlayer; the tungsten- and oxygen-originated spots, as well as the corresponding unit cells are coded as gray and orange, respectively. However, the behavior of the spot profiles as a function of energy, as shown in Figure 7b, indicates that the 113-phase represents a more complicated case with respect the 337-phase. For the 113-phase most of the intensity is shared not only between the two spots representing the tungsten and oxygen lattices; the second order satellites also reveal an increased intensity for most energies. Moreover, the intensity relationships between the spots are not monotonic, as was for the 337-phase, and there are energy ranges for which the second order satellites dominate other superstructure spots (e.g., 48-51, 70-77, and above 86 eV), as illustrated in Figure 7c. This suggests an extra periodicity in the stripe structure corresponding to the half width of the basic stripe. As demonstrated below, the proposed model can reproduce such a periodicity.

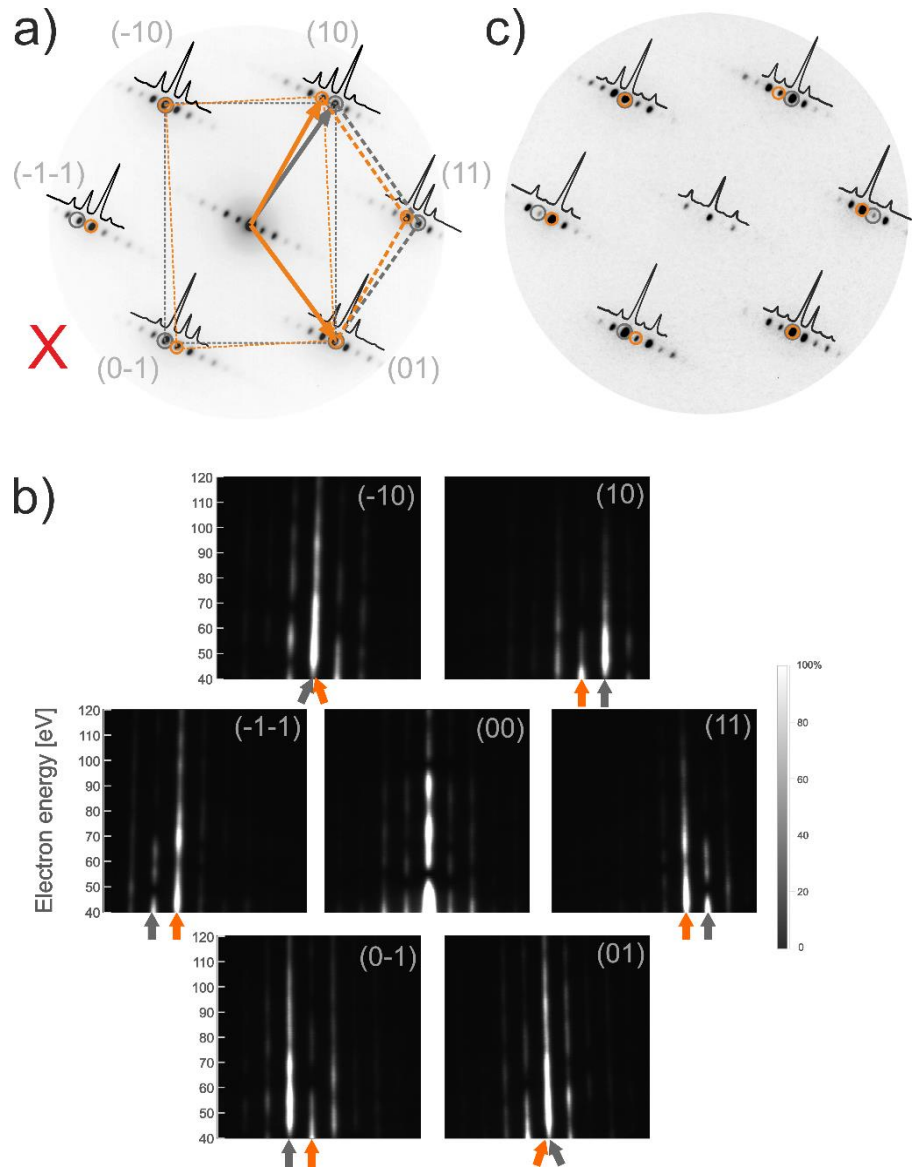

Figure 7. Data derived from the LEED patterns collected for the 113-phase X structure. a) LEED pattern with intensity profiles across the W(1x1) spots and their satellites averaged in energy over 40–70 eV. Gray and orange circles and vectors mark the spots and unit cells corresponding to tungsten and oxygen lattices, respectively. b) Changes in the intensity profiles across the superstructure arrays as functions of energy (40–120 eV). c) An example LEED pattern (collected at 98 eV) manifesting the influence of periodicity corresponding to half the stripe width.

In an earlier paper, we determined the superstructure cell of the 113-phase based on the geometry of the diffraction pattern [14]. This supercell is the result of a coincidence between an oxygen overlayer and the tungsten lattice as:

$\mathbf{G_{coinc}} = \begin{bmatrix} 10 & 1 \\ -1 & 2 \end{bmatrix}$ with the oxygen lattice given by $G_O = \begin{bmatrix} 1 & \frac{1}{10} \\ 0 & \frac{21}{20} \end{bmatrix}$, for the X domain $\mathbf{G_{coinc}} = \begin{bmatrix} 2 & -1 \\ 1 & 10 \end{bmatrix}$ with the oxygen lattice given by $\mathbf{G_O} = \begin{bmatrix} \frac{21}{20} & 0 \\ \frac{1}{10} & 1 \end{bmatrix}$, for the Y domain

One of these coincidence structures (X) is visualized in the Supplementary Information in Figure S1b. The coincidence unit cell contains 21 W vs 10x2=20 O atoms, which means that there are 4.76% fewer oxygen than tungsten atoms and 0.51% less oxygen than for the 337-phase. The $\mathbf{O_\perp}$ close-packed rows are rotated by 5.21° against the close-packed tungsten $\langle 1\bar{1}1 \rangle$ rows, and the distance between atoms in these rows is 1.0376u. For the second oxygen close-packed direction ($\mathbf{O_{II}}$), the periodicity of the oxygen atoms is 1.05u. The structural parameters of the oxygen unit cells for both phases are compared in Table 1. The superstructure for the 113-phase is a result of the same type of unidirectional reorganization within and between the oxygen $\mathbf{O_{II}}$ chains as was for the 337-phase. However, within a chain, 20 oxygen atoms are matched to 21 tungsten units indicating that these chains are more oxygen depleted compared to the 337-phase, whereas the slip between neighboring chains amounts to (1/10)u, which is less than in the 337-phase. The periodicity between chains again remains unchanged and is the same as the spacings for the close-packed tungsten rows. This chain reorganization produces moiré stripes running along one of the two mirror $\langle 1\bar{1}3 \rangle$ directions. The two possible domains corresponding to the average adsorption position at the $\overline{1b}$ site are shown in Figure 8. It is noted that for such adsorption position, the $\mathbf{O_{II}}$ chains are centered between corresponding close-packed W$\langle 1\bar{1}1 \rangle$ rows, which seems to provide the best match for the two lattices (details are included in the Supplementary Information in Figure S3c).

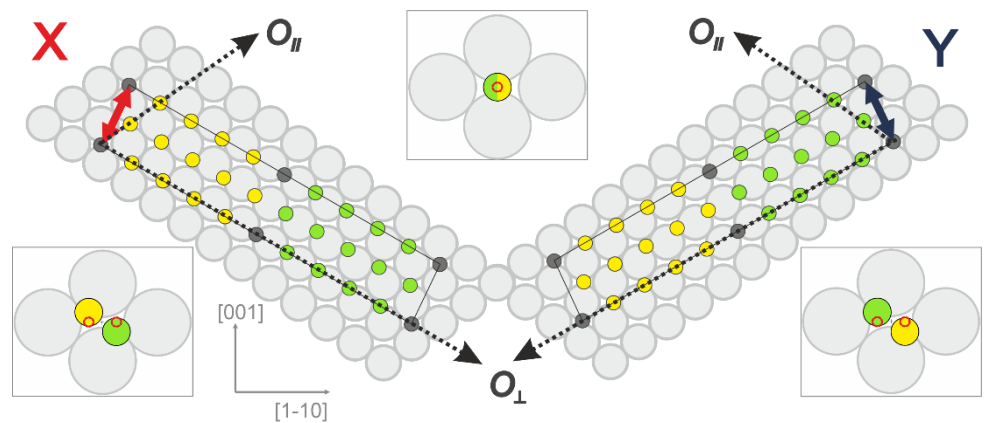

Figure 8. Coincidence unit cells for the X and Y domains in the 113-phase with an assumed $\overline{1b}$ adsorption position. The large circles represent the (110) tungsten surface, and the small circles correspond to oxygen lattice (described by one of the $\mathbf{G_O}$ matrixes in the text). The W and O lattices create two coincidence structures (described by $\mathbf{G_{coinc}}$ matrixes) that have different stripe directions (red and blue bidirectional arrows, respectively). The $\mathbf{O_\perp}$ and $\mathbf{O_{II}}$ correspond to the dense-packed rows of oxygen. Various colors for the oxygen atoms indicate different adsorption site types: yellow and green mark oxygen in left and right h3-like hollows, respectively, and dark gray are for long or short bridge positions. The pictograms in the bottom insets show the average adsorption positions within the yellow and green halves of the coincidence unit cell: the horizontal positions correspond exactly to the h3 sites. Thus, they are positioned 0.569 Å left or right from the 1b point with a 0.396-Å vertical displacement. The central inset shows the average adsorption position for the entire coincidence unit cell.

The explanation for the additional periodicity observed in the LEED patterns for the 113-phase can be rationalized based on the above model. The periodicity follows from the twofold rotational symmetry of the coincidence unit cell that splits the unit in two 180°-rotated parts (marked with yellow and green in Figure 8). The average adsorption positions within the halves are mutually symmetric with respect to the $\overline{1b}$ site and close to the $\overline{h3}$, as shown in the insets of Figure 8. In this way, the basic stripes are divided into equally wide sub-stripes, for which the mean adsorption position alternates between the two pseudo-triple hollows. This model gives alternating pseudo S-E domains with a periodicity reflected in the LEED patterns, and, in this special sense, the S-E stripe concept [4] applies to the 113-phase.

In the frame of the proposed model, the presence of unoccupied rows along the $\langle 1\bar{1}3 \rangle$ direction of the 113-phase, as observed by Johnson *et al.* [4], can be rationalized. Oxygen atoms in the short bridge (sb) positions and in their neighborhood, occupy energetically unfavorable adsorption sites; the ab initio DFT calculations indicate the adsorption energy in the sb position is approximately 0.5–0.6 eV higher than the h3 position [31,32]. The 113-phase reported by Johnson *et al.* [4] was obtained from a UHV high-temperature treatment (at ≥ 1000 ºC), during which a gradual broadening of the unoccupied rows was observed as the annealing temperature increased. Therefore, it is plausible that the least bound $\langle 1\bar{1}3 \rangle$ oxygen chains desorb during annealing.

| Phase | Atomic spacing along $O_\perp$ [u=2.7411 Å] | Atomic spacing along $O_\parallel$ [u] | Angle between $O_\perp$ and $W\langle 1\bar{1}1 \rangle$ [°] | Angle between $O_\perp$ and $O_\parallel$ [°] | $n_O/n_W$ | Relative oxygen deficiency [%] |
|---|---|---|---|---|---|---|
| 337 | 1.0423 | 1.04(4) | 5.77 | 64.76 | 45/47 | 4.25 |
| 113 | 1.0376 | 1.0500 | 5.21 | 65.32 | 20/21 | 4.76 |

Table 1. Juxtaposition of structural parameters for the oxygen unit cells for the 337- and 113-phases. The oxygen deficiency is related to the (1x1) structure.

## 4. Conclusions

We conducted a thorough analysis of the structural domains observed in the high-temperature 337- and 113-phases of an oxygen monolayer on W(110). Symmetry considerations based on our df-LEEM and μ-LEED I-V measurements and from previous df-XPEEM data [13] allowed us to shed new light on the nature of the striped superstructure, the size and morphology of the S-E domains, and the oxygen adsorption sites.

In particular, we demonstrated that the S-E domains observed in the 337-phase are at the micrometer scale, indicating the stripes do not originate from alternating S-E domains as had been commonly accepted for nearly three decades. Our analysis strongly supports an alternative explanation that the stripes stem from a misfit between the tungsten and oxygen lattices, which leads to a coincidence moiré superstructure. In our models for the 337- and 133-phases, the system relaxes the stress related to the presence of the oxygen overlayer through uniaxial expansion of the close-packed oxygen chains which remain parallel to $W\langle 1\bar{1}1 \rangle$. We showed that the oxygen density in these close-packed chains for the 337-phase is lower than in the (1x1) oxygen monolayer, and is further reduced in the 113-phase. We postulate that the uniaxial oxygen depletion in the close-packed chains is overall responsible for the observed thermal evolution of an adsorbed oxygen monolayer, which starts from the (1x1) structure at ambient temperatures and continues with increasing temperatures through the 337- up to the 113-phase. It is noted that the mechanism for the uniaxial reorganization of the oxygen overlayer (via desorption/adsorption) also poses an explanation for the experimentally observed nucleation of new phases in the form of needle-like structures that spread just along the $W\langle 1\bar{1}1 \rangle$ directions during the 113 ↔ 337 phase transitions [14].

Our scrutiny of the 337-phase confirms that parallelly to S-E effect, which breaks the mirror symmetry with respect to the [001] axis of the W(110) substrate, additional symmetry breaking with respect to [1$\bar{1}$0] axis is present. It is due to a shift in the average oxygen adsorption position from the ideal $\bar{h}$3 along one of the four equivalent non-cardinal directions. For given sub-domain the shift is unique, and is strictly correlated with left or right adsorption hollow and specific direction of the stripes. This correlation is a natural consequence of the properties for the proposed coincidence structure with the h3-type average adsorption site. As shown, to optimally match the uniaxially depleted oxygen chains to the tungsten substrate, a unique shift of the oxygen overlayer from the ideal $\bar{h}$3 position is required.

We showed that the large micro-sized S-E domains and other accompanying manifestations of the broken mirror symmetries observed for the 337-phase in the df-LEEM and LEED measurements disappear when the oxygen overlayer transforms to the 113-phase. This implies that the average adsorption position of oxygen, localized at left or right h3 hollow for the oxygen-rich 337-overlayer, moves to the center of the tungsten unit cell when the density of the oxygen monolayer is reduced in the 113-phase.

Although the presented models require further confirmation through theoretical calculations, the fact that they supply mechanisms consistently explaining all the complicated architectures of the oxides advocates their feasibility. We believe these data will stimulate theoretical investigations in this matter. We also believe that our findings will help further the development of phase diagrams for oxygen adsorption on W(110), which poses one of the oldest but still relevant classical fields in surface science.

**Acknowledgments**

We would like to thank Tevfik Onur Menteş and Andrea Locatelli, who were the first to observe the subdomains in the 337-phase, for fruitful discussions, valuable suggestions, and critical reading of the text.

This work was conducted under the statutory research funds of ICSC PAS within the subsidy of the Ministry of Science and Higher Education, Poland.

[1]     L.H. Germer, J.W. May, Diffraction study of oxygen adsorption on a (110) tungsten face, Surf. Sci. 4 (1966) 452–470. https://doi.org/10.1016/0039-6028(66)90019-7.

[2]     E. Bauer, Multiple scattering versus superstructures in low energy electron diffraction, Surf. Sci. 7 (1967) 351–354. https://doi.org/10.1016/0039-6028(67)90026-X.

[3]     E. Bauer, T. Engel, Adsorption of oxygen on W(110): II. The high coverage range, Surf. Sci. 71 (1978) 695–718. https://doi.org/10.1016/0039-6028(78)90456-9.

[4]     K.E. Johnson, R.J. Wilson, S. Chiang, Effects of adsorption site and surface stress on ordered structures of oxygen adsorbed on W(110), Phys. Rev. Lett. 71 (1993) 1055–1058.

https://doi.org/10.1103/PhysRevLett.71.1055.

[5] C.H.F. Peden, N.D. Shinn, Oxidation of W(110): valence-band and W(4f) core-level spectroscopy, Surf. Sci. 312 (1994) 151–156. https://doi.org/10.1016/0039-6028(94)90812-5.

[6] H. Daimon, R. Ynzunza, J. Palomares, H. Takabi, C.S. Fadley, Direct structure analysis of W(110)-(1 × 1)-O by full solid-angle X-ray photoelectron diffraction with chemical-state resolution, Surf. Sci. 408 (1998) 260–267. https://doi.org/10.1016/S0039-6028(98)00249-0.

[7] D.M. Riffe, G.K. Wertheim, Submonolayer oxidation of W(110): A high-resolution core-level photoemission study, Surf. Sci. 399 (1998) 248–263. https://doi.org/10.1016/S0039-6028(97)00824-8.

[8] H. Daimon, R.X. Ynzunza, F.J. Palomares, E.D. Tober, Z.X. Wang, A.P. Kaduwela, M.A. Van Hove, C.S. Fadley, Circular dichroism in core-level emission from O/W(110): Experiment and theory, Phys. Rev. B. 58 (1998) 9662–9665. https://doi.org/10.1103/PhysRevB.58.9662.

[9] R.X. Ynzunza, F.J. Palomares, E.D. Tober, Z. Wang, J. Morais, R. Denecke, H. Daimon, Y. Chen, Z. Hussain, M.A. Van Hove, C.S. Fadley, Structure determination for saturated (1 × 1) oxygen on W(110) from full solid angle photoelectron diffraction with chemical-state resolution, Surf. Sci. 442 (1999) 27–35. https://doi.org/10.1016/S0039-6028(99)00787-6.

[10] J. Feydt, A. Elbe, H. Engelhard, G. Meister, A. Goldmann, Normal-emission photoelectron studies of the W(110)(1 × 1)O surface, Surf. Sci. 440 (1999) 213–220. https://doi.org/10.1016/S0039-6028(99)00796-7.

[11] H. Takagi, H. Daimon, F.J. Palomares, C.S. Fadley, Photoelectron holography analysis of W(1 1 0)(1×1)-O surface, Surf. Sci. 470 (2001) 189–196. https://doi.org/10.1016/S0039-6028(00)00839-6.

[12] K. Radican, S.I. Bozhko, S.R. Vadapoo, S. Ulucan, H.C. Wu, A. McCoy, I. V. Shvets, Oxidation of W(110) studied by LEED and STM, Surf. Sci. 604 (2010) 1548–1551. https://doi.org/10.1016/j.susc.2010.05.016.

[13] T.O. Menteş, A. Locatelli, Angle-resolved X-ray photoemission electron microscopy, J. Electron Spectros. Relat. Phenomena. 185 (2012) 323–329. https://doi.org/10.1016/j.elspec.2012.07.007.

[14] T. Giela, D. Wilgocka-Ślęzak, M. Ślęzak, N. Spiridis, J. Korecki, LEEM study of high-temperature oxygen structures on W(110) and their transformations, Appl. Surf. Sci. 425 (2017) 314-320. https://doi.org/10.1016/j.apsusc.2017.07.020.

[15] R. Gomer, J.K. Hulm, Adsorption and diffusion of oxygen on tungsten, J. Chem. Phys. 27 (1957) 1363–1376. https://doi.org/10.1063/1.1744008.

[16] R. Butz, H. Wagner, Diffusion of oxygen on tungsten (110), Surf. Sci. 63 (1977) 448–459. https://doi.org/10.1016/0039-6028(77)90358-2.

[17] J.-R. Chen, R. Gomer, Mobility of oxygen on the (110) plane of tungsten, Surf. Sci. 79 (1979) 413–444. https://doi.org/10.1016/0039-6028(79)90298-X.

[18] M. Tringides, R. Gomer, Anisotropy in surface diffusion: Oxygen, hydrogen, and deuterium on the (110) plane of tungsten, Surf. Sci. 155 (1985) 254–278. https://doi.org/10.1016/0039-6028(85)90417-0.

[19]   T.U. Nahm, R. Gomer, The diffusion of oxygen on W(110) revisited, J. Chem. Phys. 106 (1997) 10349–10358. https://doi.org/10.1063/1.474104.

[20]   C. Uebing, R. Gomer, The diffusion of oxygen on W(110) The influence of the p(2 × 1) ordering, Surf. Sci. 381 (1997) 33–48. https://doi.org/10.1016/S0039-6028(97)00081-2.

[21]   I. Vattulainen, J. Merikoski, T. Ala-Nissila, Adatom dynamics and diffusion in a model of O/W(110), Phys. Rev. B  57 (1998) 1896–1907. https://doi.org/10.1103/PhysRevB.57.1896.

[22]   P.K. Wu, M.C. Tringides, M.G. Lagally, Ordering kinetics of a chemisorbed overlayer: O/W(110), Phys. Rev. B. 39 (1989) 7595–7610. https://doi.org/10.1103/PhysRevB.39.7595.

[23]   M.C. Tringides, Growth kinetics of O/W(110) at high coverage, Phys. Rev. Lett. 65 (1990) 1372–1375. https://doi.org/10.1103/PhysRevLett.65.1372.

[24]   R.X. Ynzunza, R. Denecke, F.J. Palomares, J. Morais, E.D. Tober, Z. Wang, F.J. García De Abajo, J. Liesegang, Z. Hussain, M.A. Van Hove, C.S. Fadley, Kinetics and atomic structure of O adsorption on W(110) from time- and state-resolved photoelectron spectroscopy and full-solid-angle photoelectron diffraction, Surf. Sci. 459 (2000) 69–92. https://doi.org/10.1016/S0039-6028(00)00450-7.

[25]   G. Ertl, D. Schillinger, Interactions between chemisorbed atoms: Oxygen on W(110), J. Chem. Phys. 66 (1977) 2569–2573. https://doi.org/10.1063/1.434254.

[26]   E.D. Williams, S.L. Cunningham, W.H. Weinberg, A determination of adatom-adatom interaction energies: Application to oxygen chemisorbed on the tungsten (110) surface, J. Chem. Phys. 68 (1978) 4688–4693. https://doi.org/10.1063/1.435579.

[27]   J.C. Buchholz, M.G. Lagally, Order-Disorder Transition and Adatom-Adatom Interactions in a Chemisorbed Overlayer: Oxygen on W(110), Phys. Rev. Lett. 35 (1975) 442–445. https://doi.org/10.1103/PhysRevLett.35.442.

[28]   T.-M. Lu, G.-C. Wang, M.G. Lagally, Island-Dissolution Phase Transition in a Chemisorbed Layer, Phys. Rev. Lett. 39 (1977) 411–414. https://doi.org/10.1103/PhysRevLett.39.411.

[29]   G.-C. Wang, T.-M. Lu, M.G. Lagally, Phase transitions in the chemisorbed layer W(110) p (2×1)–O as a function of coverage. I. Experimental, J. Chem. Phys. 69 (1978) 479. https://doi.org/10.1063/1.436377.

[30]   D.H. Baek, J.W. Chung, W.K. Han, Critical behavior of the p(2×1)-O/W(110) system, Phys. Rev. B. 47 (1993) 8461–8464. https://doi.org/10.1103/PhysRevB.47.8461.

[31]   M.A. Załuska-Kotur, S. Krukowski, Z. Romanowski, Ł.A. Turski, Twin spin model of surface phase transitions in O/W(110), Phys. Rev. B  65 (045404) (2001) 1-9. https://doi.org/10.1103/PhysRevB.65.045404.

[32]   M. Stöhr, R. Podloucky, S. Müller, Ab initio phase diagram of oxygen adsorption on W(110), J. Phys. Condens. Matter. 21 (2009) 134017 (10pp). https://doi.org/10.1088/0953-8984/21/13/134017.

[33]   T. Engel, H. Niehus, E. Bauer, Adsorption of oxygen on W(110), Surf. Sci. 52 (1975) 237–262. https://doi.org/10.1016/0039-6028(75)90056-4.

[34]   M.A. Van Hove, S.Y. Tong, Chemisorption bond length and binding location of oxygen in a p(2×1) overlayer on W(110) using a convergent, perturbative, low-energy-electron-diffraction

calculation, Phys. Rev. Lett. 35 (1975) 1092–1095. https://doi.org/10.1103/PhysRevLett.35.1092.

[35] J.M. Baker, D.E. Eastman, Photoemission spectra from adsorbed O on W(110) and CO on W(100), J. Vac. Sci. Technol. 10 (1973) 223–226. https://doi.org/10.1116/1.1317946.

[36] N.J. Dinardo, G.B. Blanchet, E.W. Plummer, Vibrational modes of oxygen on W(110), Surf. Sci. 140 (1984) L229–L238. https://doi.org/10.1016/0039-6028(84)90373-X.

[37] A. Elbe, G. Meister, A. Goldmann, Vibrational modes of atomic oxygen on W(110), Surf. Sci. 371 (1997) 438–444. https://doi.org/10.1016/S0039-6028(96)10096-0.

[38] M.J. Harrison, D.P. Woodruff, J. Robinson, Density functional theory calculations of adsorption-induced surface stress changes, Surf. Sci. 602 (2008) 226–234. https://doi.org/10.1016/j.susc.2007.10.011.

[39] T.O. Menteş, N. Stojić, N. Binggeli, M.A. Niño, A. Locatelli, L. Aballe, M. Kiskinova, E. Bauer, Strain relaxation in small adsorbate islands: O on W(110), Phys. Rev. B 77 (155414) (2008) 1-9. https://doi.org/10.1103/PhysRevB.77.155414.

[40] N. Stojić, T.O. Menteş, N. Binggeli, M.A. Niño, A. Locatelli, E. Bauer, Temperature dependence of surface stress across an order-disorder transition: P (1×2 ) O/W (110 ), Phys. Rev. B 81 (115437) (2010) 1–7. https://doi.org/10.1103/PhysRevB.81.115437.

[41] T. Engel, T. von dem Hagen, E. Bauer, Adsorption and desorption of oxygen on stepped tungsten surfaces, Surf. Sci. 62 (1977) 361–378. https://doi.org/10.1016/0039-6028(77)90088-7.

[42] T.E. Madey, The role of steps and defects in electron stimulated desorption: Oxygen on stepped W(110) surfaces, Surf. Sci. 94 (1980) 483–506. https://doi.org/10.1016/0039-6028(80)90021-7.

[43] M. Mašín, I. Vattulainen, T. Ala-Nissila, Z. Chvoj, Interplay between steps and nonequilibrium effects in surface diffusion for a lattice-gas model of O/W(110), J. Chem. Phys. 126 (2007) 114705. https://doi.org/10.1063/1.2713100.

[44] H. Zheng, J.Z. Ou, M.S. Strano, R.B. Kaner, A. Mitchell, K. Kalantar-Zadeh, Nanostructured tungsten oxide - Properties, synthesis, and applications, Adv. Funct. Mater. 21 (2011) 2175–2196. https://doi.org/10.1002/adfm.201002477.

**SUPPLEMENTARY INFORMATION**

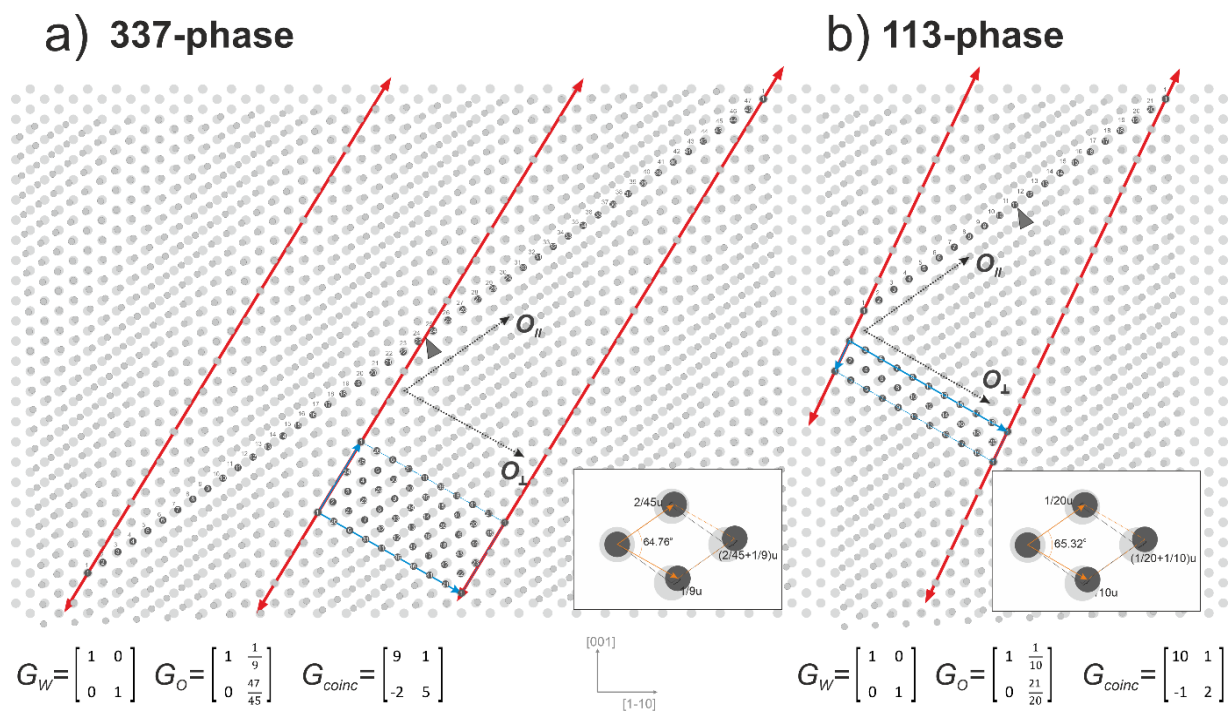

## a) 337-phase

## b) 113-phase

$$G_W = \begin{bmatrix} 1 & 0 \\ 0 & 1 \end{bmatrix} \quad G_O = \begin{bmatrix} 1 & \frac{1}{9} \\ 0 & \frac{47}{45} \end{bmatrix} \quad G_{coinc} = \begin{bmatrix} 9 & 1 \\ -2 & 5 \end{bmatrix}$$

$$G_W = \begin{bmatrix} 1 & 0 \\ 0 & 1 \end{bmatrix} \quad G_O = \begin{bmatrix} 1 & \frac{1}{10} \\ 0 & \frac{21}{20} \end{bmatrix} \quad G_{coinc} = \begin{bmatrix} 10 & 1 \\ -1 & 2 \end{bmatrix}$$

Figure S1. Visualization of coincidence structures for the a) 337-phase (corresponding to A or B sub-domains) and the b) 113-phase (X domain). Larger light-gray circles depict tungsten whereas the smaller darker-gray circles represent oxygen atoms. The red bidirectional arrows indicate the direction of the stripes and their periodicity, the blue vectors mark the coincidence unit cells, and the black dashed vectors point from the $O_\perp$ and $O_\parallel$ close-packed oxygen chains. To highlight the most characteristic features of the coincidence between the oxygen and tungsten lattices, the adsorption positions were assumed arbitrarily as being on-top. Note that deformation of the oxygen lattice with respect to the tungsten structure in these specific domains occur via loss of oxygen atoms from the individual [1$\bar{1}$1] chains and by mutual sliding of these chains along this direction. An oxygen chain corresponding to a single coincidence unit cell is accentuated with darker gray shades. (The O and W atoms are numbered by white and gray numbers, respectively. For comparison, the same numeration is applied for oxygen atoms within the rhomboidal coincidence unit cell.) The gray triangles mark the centers of the chains. INSET: W and O unit cells with values corresponding to the displacements of the oxygen atoms with respect to tungsten.

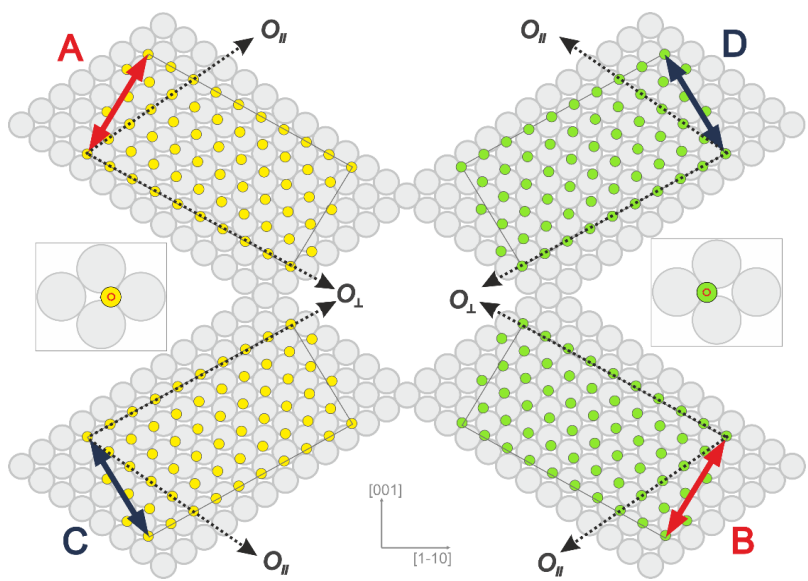

Figure S2. Coincidence unit cells for the A, B, C, and D sub-domains in the 337-phase with an average ideal $\bar{h}3$ position assumed. Larger circles represent the (110) tungsten surface, while smaller circles correspond to the oxygen lattice (described by the $G_O$ matrices as seen in the text). The W and O lattices create two types of coincidence structures (described with $G_{coinc}$ matrices) that differ in the direction of the stripes (red and blue bidirectional arrows). However, as the average adsorption site in this structure is $\bar{h}3$ and there are two sites (coded as green and yellow), the number of possible domains increases to four. The $O_\perp$ and $O_\parallel$ correspond to close-packed oxygen rows. Pictograms in the insets show the average adsorption positions within the coincidence unit cells (small red ring marks the exact h3 position). Note that this structure gives no contrast in the df-LEEM performed at the (1-1) and (-11) spots due to the conserved symmetry of the average adsorption site along the [001] direction.

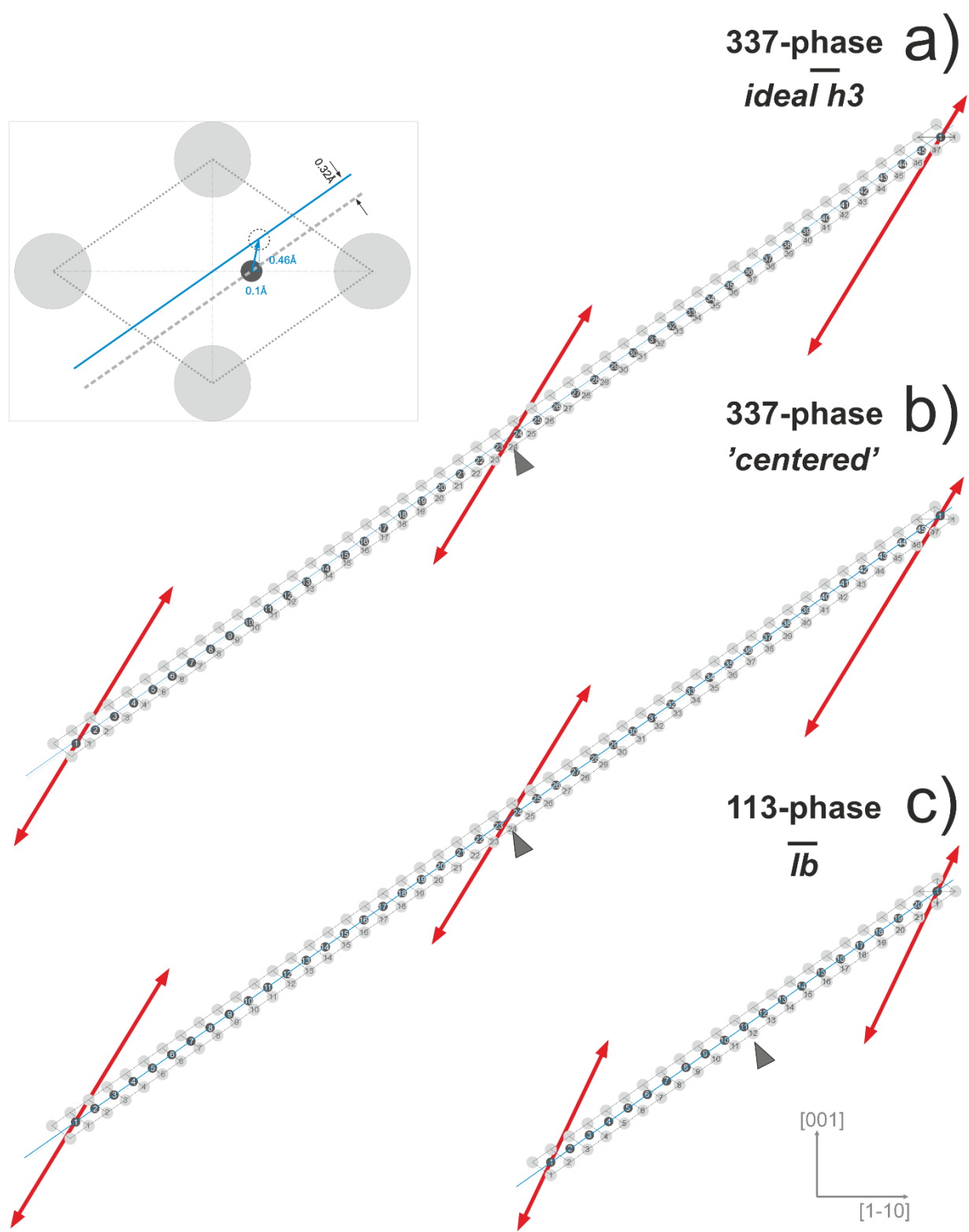

Figure S3. Visualization of changes in the local adsorption position within the coincidence unit cell represented as an **O**$_{III}$ chain (darker circles) across the tungsten unit cells (lighter circles) for the 337-phase with the A sub-domain for oxygen, on average, at the a) ideal $\overline{h3}$ position and b) **O**$_{III}$ centered along tungsten [1$\overline{1}$1] groove, for c) the 113-phase with the X domain with oxygen, on average, at the long bridge site ($\overline{lb}$). Note that for b) and c), the oxygen chains lay ideally between the tungsten close-packed [$\overline{1}$11] chains (the center of the groove is marked with a blue line) whereas for the 'ideal h3,' the oxygen chain is localized closer to the bottom tungsten chain (marked with a dashed gray line). The red bidirectional arrow marks the direction of the stripes and their periodicities. The gray arrows point to the center of the superstructures. The inset compares the average positions of the oxygen for the ideal $\overline{h3}$ position (dark circle) and for the 'groove' model (circle marked with a dashed-line).